\documentclass[sn-standardnature]{sn-jnl}

\usepackage[resetlabels,labeled]{multibib}
\newcites{S}{Supplementary Material References}
\usepackage{longtable}
\usepackage{pdflscape}
\usepackage{dirtytalk}
\usepackage{microtype}
\DeclareUnicodeCharacter{2009}{ }
\jyear{2023}

\begin{document}

\title[Gender Inclusive Methods in Studies of STEM Practitioners]{Gender Inclusive Methods in Studies of STEM Practitioners}

 \author[1]{\fnm{Kaitlin} \sur{Rasmussen}}\email{Kaitlin Rasmussen}
 \author[2]{\fnm{Jocelyne} \sur{Chen}}\email{Jocelyne Chen}
 \author*[3]{\fnm{Rebecca L.} \sur{Colquhoun}}\email{rebeccalcolquhoun@gmail.com}
 \author[4]{\fnm{Sophia} \sur{Frentz}}\email{Sophia Frentz}
 \author[5]{\fnm{Laurel} \sur{Hiatt}}\email{Laurel Liatt}
 \author[6,7]{\fnm{Aiden James} \sur{Kosciesza}}\email{Aiden James Kosciesza}
 \author[8]{\fnm{Charlotte} \sur{Olsen}}\email{Charlotte Olsen}
 \author[9]{\fnm{Theo J.} \sur{O'Neill}}\email{Theo J. O'Neill}
 \author[10]{\fnm{Vic} \sur{Zamloot}}\email{Vic Zamloot}
 \author{\fnm{Beckett E.} \sur{Strauss}}\email{Beckett E. Strauss}

\affil[1]{\orgdiv{Department of Science, Engineering, and Mathematics}, \orgname{Tacoma Community College}, \orgaddress{\city{Tacoma}, \postcode{98466,}, \state{WA}, \country{USA}}}
\affil[2]{\orgname{The New School for Social Research}, \orgaddress{\street{66 West 12th Street}, \city{New York}, \postcode{10011}, \state{NY}, \country{USA}}}
 \affil[3]{\orgdiv{Department of Earth Sciences}, \orgname{University of Oxford}, \orgaddress{\street{3 South Parks Road}, \city{Oxford}, \postcode{OX1 3AN}, \country{United Kingdom}}}
 \affil[4]{\orgname{Royal Society of Victoria}, \orgaddress{\street{8 La Trobe Street}, \city{Melbourne}, \postcode{3000}, \state{Victoria}, \country{Australia}}}
\affil[5]{\orgdiv{School of Medicine}, \orgname{University of Utah}, \orgaddress{\street{30 N 1900 E}, \city{Salt Lake City}, \postcode{84132}, \state{Utah}, \country{USA}}}
 \affil[6]{\orgname{Temple University}, \orgaddress{\street{2020 N 13th St}, \city{Philadelphia}, \postcode{19122}, \state{PA}, \country{USA}}}
 \affil[7]{ \orgname{Community College of Philadelphia}, \orgaddress{\street{1700 Spring Garden St}, \city{Philadelphia}, \postcode{19130}, \state{PA}, \country{USA}}}
\affil[8]{\orgname{Rutgers University}, \orgaddress{\street{136 Frelinghuysen Rd}, \city{Piscataway}, \postcode{08854}, \state{NJ}, \country{USA}}}
\affil[9]{\orgdiv{Center for Astrophysics}, \orgname{Harvard \& Smithsonian}, \orgaddress{\street{60 Garden St.}, \city{Cambridge}, \postcode{02138}, \state{MA}, \country{USA}}}
 \affil[10]{\orgdiv{Irell and Manella Graduate School of Biological Sciences}, \orgname{City of Hope}, \orgaddress{\street{1218 S 5th Ave}, \city{Monrovia}, \postcode{91016}, \state{CA}, \country{USA}}}

\abstract{Gender inequity is one of the biggest challenges facing the STEM workforce. While there are many studies that look into gender disparities within STEM and academia \cite{corbett_solving_2015,dean_equitable_2014,diele-viegas_community_2022,hill_why_2010,mcmahon_advancing_2018,noonan_women_2017}, the majority of these have been designed and executed by those unfamiliar with research in sociology and gender studies. They adopt a normative view of gender as a binary choice of `male' or `female,’ leaving individuals whose genders do not fit within that model out of such research entirely \cite{kosciesza_intersectional_2022,namaste_invisible_2000}. This especially impacts those experiencing multiple axes of marginalization, such as race, disability, and socioeconomic status. For STEM fields to recruit \textit{and} retain members of historically excluded groups, a new paradigm must be developed. Here, we collate a new dataset of the methods used in 119 past studies of gender equity, and recommend better survey practices and institutional policies based on a more complex and accurate approach to gender. We find that problematic approaches to gender in surveys can be classified into 5 main themes---treating gender as white, observable, discrete, as a statistic, and as inconsequential. We recommend allowing self-reporting of gender and never automating gender assignment within research. This work identifies the key areas of development for studies of gender-based inclusion within STEM, and provides recommended solutions to support the methodological uplift required for this work to be both scientifically sound and fully inclusive.
}




\maketitle

\section{Introduction}
The last several decades have seen significant advances in the inclusion of women \cite{NCSES, IPEDS} in STEM, and countless studies, community efforts, and institutional initiatives have focused on addressing their status
\cite{corbett_solving_2015,dean_equitable_2014,diele-viegas_community_2022,hill_why_2010,mcmahon_advancing_2018,noonan_women_2017}. More recently, transgender (i.e., people whose gender identity, expression, or behavior does not conform to that typically associated with the sex to which they were assigned at birth) and nonbinary (an umbrella term for all genders not represented by the categories of ‘male’ or ‘female’ \cite{rasmussen_nonbinary_2019}) identities have also gained recognition and inclusion in STEM equity efforts \cite{armada-moreira_stem_2021,cross_queering_2022,unsay_lgbtq_2020}, with the rise of many professional societies \cite{500qs, ISNBS, QiSAus} aimed at reducing institutional barriers for LGBTQ2S+ scientists.


However, many gender-related equity initiatives and studies have been led by professional STEM practitioners with minimal background or training in social sciences or gender studies. As a result, these studies often use methodologies that range from overly simplistic to actively harmful. In this work, we first summarize recent work on gender equity in the various sub-fields of STEM. We then discuss common pitfalls of such work, with a focus on their detrimental effects on the inclusion of transgender and nonbinary STEM practitioners. Finally, we offer recommendations for studying gender dynamics and promoting gender inclusion in STEM going forward. 

\section{Previous Studies}

Numerous studies have attempted to evaluate and address gender disparities across all STEM fields and within particular disciplines, including (but not limited to) the geosciences, physics, astronomy, planetary science, biology, engineering, and chemistry. Most of these publications were written by professional STEM researchers, and all were intended for and circulated among audiences composed of STEM practitioners. While the literature review presented here (Table \ref{table:1}) is far from comprehensive, it is broadly illustrative of the concepts and methods that STEM researchers employ in studying gender-related phenomena within their fields.

In STEM, these studies most commonly examine the impact of gender on career-related metrics. They may examine first-author publication rates, resource allocation, research team gender diversity, time between PhD completion and being hired into a long-term position, and more. Another topic of focus is social dynamics in professional settings, such as harassment and discrimination with regards to gender and sexuality, or conference participation as a function of the gender of the question-asker, the gender of the speaker, and the gender of the session chair.

\section{Problematic Methods}

We build on the prior work of Rasmussen et al. (2019) \cite{rasmussen_nonbinary_2019} and Strauss et al. (2020) \cite{strauss_nonbinary_2020} and update the framing of this work with respect to a key concern: the intersectionality of race and gender. These previous works identified four major concerns common to most of the endeavors described in Table \ref{table:1}: ``{[(2)]} the treatment of gender as observable through means other than self-identification; {[(3)]} categorization schemes with limited gender options; {[(4)]} an over-reliance on quantitative methodology that is at best insufficient for understanding gendered phenomena in STEM and at worst epistemically violent towards people whose genders are poorly represented by these schema; {[(5)]} the treatment of nonbinary and transgender identities as inconsequential and therefore dismissable.'' To these we add another: (1) the treatment of gender as white, i.e., disregard for the racialized construction of gender. 

Studies of gender equity performed by STEM practitioners demonstrate little, if any, engagement with the enormous bodies of existing work in relevant disciplines including gender studies, transgender studies, sociology, and public health. They further fail to prioritize the participatory inclusion and testimony of marginalized people, often discarding data on nonbinary people even when it is collected. These methodological choices negatively impact people of all marginalized genders \cite{kosciesza_intersectional_2022,namaste_invisible_2000}. In failing to account for nonbinary people, many studies undermine their stated goal of promoting gender equity in STEM.

\newpage
\clearpage
\begin{landscape}

\textbf{Table 1:} Selected recent works on gender equity within STEM in alphabetical order by first author, with primary subject and method(s) of determining gender. Full citations are provided below.\\
\begin{longtable}{ p{.5\textwidth} | p{.2\textwidth} |p{.75\textwidth} }

Citation              & Subject    & Method                          
\\\hline \endhead
Adams \& Morgan (2021) \cite{adams_relational_2021}& Technology & No methods provided \\
Alegria (2019) \cite{alegria_escalator_2019} & Technology & No methods provided             \\
Alfrey \& Twine (2017) \cite{alfrey_gender-fluid_2017} & Technology & Asked, allowing diverse answers \\
Aycock et al. (2019) \cite{aycock_sexual_2019} & Physics &  Asked, allowing diverse answers; Excluded nonbinary respondents\\
Banning et al. (2007) \cite{banning_gender_2007} & Geoscience & Photographs (manual) \\
Benjamens et al. (2020) \cite{benjamens_gender_2020} & Medical Sciences & First names (algorithm) \\
Blair-Loy et al. (2017) \cite{blair-loy_gender_2017} & Engineering & No methods provided \\
Borsuk et al. (2009) \cite{borsuk_influence_2009} & Ecology & First names \\
Budden et al. (2008) \cite{budden_double-blind_2008} & Ecology & First names \\
Campbell et al. (2013) \cite{campbell_gender-heterogeneous_2013}& Ecology    & First names; photos   \\
Campbell et al.	(2019) \cite{campbell_authorship_2019} & Medical Sciences & No methods provided \\
Caplar et al. (2017) \cite{caplar_quantitative_2017}	& Astronomy & First names (algorithm) \\
Carr et al. (2019) \cite{carr_gender_2018} &	Medical Sciences & Asked, unclear options\\
Cech \& Blair-Loy (2010) \cite{cech_perceiving_2010} & Technology & No methods provided\\
Cech \& Rothwell (2018) \cite{cech_lgbtq_2018} & Engineering & Asked, allowing diverse answers;  Men + nonbinary grouped \\
Cech \& Pham (2017) \cite{cech_queer_2017} & STEM-wide & Asked, M/F only \\
Centrella et al. (2019) \cite{centrella_leadership_2019}	& Astronomy & First names (algorithm)\\
Clancy et al. (2014) \cite{clancy_survey_2014} &	Biology & Asked, allowing diverse answers; Excluded nonbinary respondents \\
Clancy et al. (2017) \cite{clancy_double_2017}&	Astronomy & Asked, allowing diverse answers; Excluded nonbinary respondents\\
Concannon \& Barrow (2009) \cite{concannon_cross-sectional_2009}& Engineering & Asked, M/F only\\
Cotton \& Seiple (2020) \cite{cotton_examining_2020} & Chemistry & First names (algorithm) \\
Dai \& Peterson (2020) \cite{dai_characteristics_2020} & Medical Sciences & Used American Medical Association Physician Masterfile \\
Daniels et al. (2019) \cite{daniels_navigating_2019} & STEM-wide & No methods provided \\
Davenport (2014) \cite{davenport_studying_2014} & Astronomy & Live eyes  \\
De Rosa et al. (2019) \cite{de_rosa_increasing_2019} & Astronomy & First names; Photos/videos (manual review of gender expression) \\
Delaney \& Devereux \cite{delaney_understanding_2019} & STEM-wide & School database \\
de Kleijn et al. (2020) \cite{de_kleijn_researcher_2020} & STEM-wide & First names (algorithm); Asked, unclear options \\
Dresden et al. (2017) \cite{dresden_no_2018} & STEM-wide & No methods provided \\
Dutt et al. (2016) \cite{dutt_gender_2016} & Geoscience & No methods provided \\
Dworkin et al. (2019) \cite{dworkin_extent_2020} & Biology & First names (algorithm) \\
Eaton et al. (2020) \cite{eaton_how_2020} & STEM-wide & Asked, unclear options \\
Elsevier (2019) \cite{elsevier_gender_2019} & STEM-wide & First names (algorithm) \\
England et al. (2019) \cite{england_student_2019} & Biology & Asked, allowing diverse answers; Excluded nonbinary respondents \\
Engwell et al. (2020) \cite{engwell_volcanic_2020} & Geoscience & Registration data (M/F); Asked, M/F/prefer not to say options only \\
Ermert et al. (2022) \cite{ermert_2022}& Geoscience & First names (algorithm) \\
Faulkner (2009) \cite{faulkner_doing_2009} & Engineering & No methods provided \\
Feldon et al. (2020) \cite{feldon_time-credit_2017} & Biology & No methods provided \\
Fernandes et al. (2020) \cite{fernandes_enriching_2020} & Geoscience & Search text; Used personal knowledge, website information and listed pronouns \\
Fischer et al. (2021) \cite{fischer_leveraging_2021} & Geoscience & Asked, allowing diverse answers; Excluded nonbinary respondents \\
Flaherty (2018) \cite{flaherty_leaky_2018} & Astronomy & Photos/videos; Search text \\
Fouad et al. (2019) \cite{fouad_exploring_2020} & Engineering & No methods provided \\
Fournier et al. (2020) \cite{fournier_unpaid_2019} & STEM-wide & Asked, unclear options \\
Fox et al. (2016) \cite{fox_gender_2016} & Ecology & First names (algorithm); Photos/videos; Search text \\
Fox et al. (2018) \cite{fox_patterns_2018} & Ecology & First names (algorithm) \\
Fox et al. (2019) \cite{fox_gender_2019} & Ecology & First names (algorithm); Asked \\
Fox \& Paine (2019) \cite{fox_gender_2019-1} & Ecology & First names (algorithm); Asked  \\
Franklin et al. (2021) \cite{franklin_who_2021} & Geoscience & Search text \\
Freire et al. (2020) \cite{freire_measuring_2020} & Computer Science & First names (algorithm) \\
Ghiasi et al. (2015) \cite{ghiasi_compliance_2015} & Engineering  & First names \\
Gonsalves (2018) \cite{gonsalves_exploring_2018} & Astronomy  & No method provided \\
Good et al. (2012) \cite{good_why_2012} & Mathematics  & No method provided \\
Haven et al. (2019) \cite{haven_perceived_2019} & STEM-wide  & Asked, M/F only \\
Hechtman et al. (2018) \cite{hechtman_nih_2018} & Medical Sciences  & Asked, unclear options \\
Herzig (2010) \cite{herzig_women_2010} & Mathematics  & No methods provided \\
Holman \& Morandin (2015) \cite{holman_researchers_2019} & Engineering  & First names (algorithm)  \\
Huang et al. (2020) \cite{huang_historical_2020} & STEM-wide  & First names (algorithm) \\
Hughes (2018) \cite{hughes_coming_2018} & STEM-wide  & Asked, unclear options \\
Hunt et al. (2021) \cite{hunt_gender-related_2019} & Astronomy  & Photos/videos; Live eyes; Asked, M/F/prefer not to say only \\
Huppenkothen et al. (2019) \cite{huppenkothen_entrofy_2020} & Astronomy  & No methods provided \\
Isbell et al. (2012) \cite{isbell_stag_2012} & Biology & First names; Search text \\
Kalejta \& Palmenberg (2017) \cite{kalejta_gender_2017} & Molecular Biology & Live eyes; Search text \\
Kaminksi \& Geisler (2012) \cite{kaminski_survival_2012} & STEM-wide & Search text \\
Kavanagh et al. (2021) \cite{kavanagh_volcanologists_2022} & Geoscience & Membership data from organizations \\
Kewley (2021) \cite{kewley_closing_2021} & Astronomy & Reported university data, M/F only \\
Ko et al. (2013) \cite{ko_narratives_2013} & Physics & No methods provided \\
Krukowski et al. (2021) \cite{krukowski_academic_2021} & STEM-wide & Asked, allowing diverse answers; Excluded nonbinary respondents  \\
Lai et al. (2021) \cite{lai_speaking_2021} & Medical Sciences & Search text \\
Lerchenmueller et al. (2020) \cite{lerchenmueller_gender_2019} & Medical Sciences & First names (algorithm) \\
Lonsdale et al. (2016) \cite{lonsdale_gender-related_2016} & Astronomy & Search text \\
Maas et al. (2021) \cite{maas_women_2021} & Ecology & First names; Photos/videos; Search text; Asked, unclear options \\
Maule\'{o}n \& Bordons (2012) \cite{mauleon_authors_2012} & Mathematics & First names; Live eyes; Search text; Asked, unclear options \\
Mickey (2019) \cite{mickey_when_2019} & Molecular & No methods provided \\
Mihalijev\'{i}c-Brandt et al. (2016) \cite{mihaljevic-brandt_effect_2016} & Mathematics & First names (algorithm); Personal knowledge \\
Milz (2018) et al. \cite{milz_who_2018} & Computer Science & First names \\
Moratti (2020) \cite{moratti_low-openness_2020} & STEM-wide & Asked, M/F only \\
Morton (2018) \cite{morton_understanding_2018} & STEM-wide & No methods provided \\
Moss-Racusin (2012) \cite{milz_who_2018} &  STEM-wide & No methods provided \\
Nadile (2021) \cite{nadile_gender_2021} & STEM-wide & Asked, allowing diverse answers; Excluded nonbinary respondents \\
National Center for Science and Engineering Statistics (2021) \cite{national_center_for_science_and_engineering_statistics_women_2021} & STEM-wide & Asked, unclear options \\
Ni et al. (2021) \cite{ni_gendered_2021} & STEM-wide & Asked, options M/F/O \\
Nielsen et al. (2017) \cite{nielsen_one_2017} & Medical Sciences & First names (algorithm) \\
Patat (2016) \cite{patat_gender_2016} & Astronomy & First names; Search text; Personal knowledge \\
Perley (2019) \cite{perley_gender_2019} & Astronomy & First names; Photos/videos; Search text \\
Piccoli \& Guidobaldi (2021) \cite{piccoli_report_2021} & Geosciences & No method provided; data taken from attendance lists and webpages \\
Pico et al. (2020) \cite{pico_first_2020} & Geosciences & First names (algorithm) \\
Primas (2019) \cite{primas_promoting_2019} & Astronomy & Two data sources: one M/F/prefer not to say, one allowed diverse answers \\
Pritchard et al. (2014) \cite{pritchard_asking_2014} & Astronomy & Live eyes \\
Ranganathan et al. (2021) \cite{ranganathan_trends_2021} & Geosciences & Search text (pronouns) \\
Rathbun (2017) \cite{rathbun_participation_2017} & Planetary Science & No methods provided \\
Rathbun et al. (2019) \cite{rathbun_whats_2019} & Planetary Science & First names; Photos/videos; Search text \\
Reid (2014) \cite{reid_gender-correlated_2014} & Astronomy & First names; Search text \\
Richey et al. (2020) \cite{richey_gender_2020} & Astronomy & Asked, allowing diverse answers; Excluded nonbinary respondents \\
Rodr\'{i}guez et al. (2020) \cite{rodriguez_gender_2020} & Mathematics & No methods provided \\
Royal Society of Chemistry (2019) \cite{royal_society_of_chemistry_is_nodate} & Chemistry & First names (algorithm) \\
Salerno et al. (2019) \cite{salerno_male_2019} & Ecology & First names; Photos/videos \\
Sassler et al. (2017) \cite{sassler_missing_2017} & STEM-wide & Asked, M/F only \\
Schmidt \& Davenport (2017) \cite{schmidt_who_2017} & Astronomy & Live eyes \\
Schmidt et al. (2017) \cite{schmidt_role_2017} & Astronomy & First names; Photos/videos; Live eyes \\
Schroeder et al. (2013) \cite{schroeder_fewer_2013} & Biology & First names; Live eyes; Search text \\
Sebbane et al. (2022) \cite{sebbane_representation_2022} & Medical Sciences & First names; Search text \\
Severin et al. (2020) \cite{severin_gender_2020} & STEM-wide & No methods provided \\
Sexton et al. (2014) \cite{sexton_characteristics_2014} & Geoscience & Photos/videos \\
Shishkova et al. (2017) \cite{shishkova_gender_2017} & Biology & First names (algorithm); Search text \\
Simard et al. \cite{simard_climbing_nodate} & Technology & Asked, unclear options \\
Sobieraj \& Kr\"{a}mer (2019) \cite{sobieraj_impacts_2019} & STEM-wide & Asked, M/F only \\
Squazzoni et al. (2021) \cite{squazzoni_peer_2021} & STEM-wide & First names (algorithm) \\
Start \& McCauley (2020) \cite{start_gender_2020} & STEM-wide & Asked, allowing diverse answers; Excluded nonbinary respondents \\
Stokes et al. (2015) \cite{stokes_choosing_2015} & Geosciences & Asked, allowed diverse answers \\
Tendhar et al. (2017) \cite{tendhar_effects_2017} & Engineering & Asked, M/F only \\
Topaz \& Sen (2016) \cite{topaz_gender_2016} & Mathematics & First names (algorithm); Search text; Excluded nonbinary respondents \\
Twine (2018) \cite{twine_technologyaposs_2018} & Technology & No methods provided \\
Urquhart-Cronish \& Otto (2019) \cite{urquhart-cronish_gender_2019} & STEM-wide & Used data from Natural Sciences and Engineering Research Council (M/F/not indicated) \\
Vila-Concejo et al. (2018) \cite{vila-concejo_steps_2018} & Geosciences & First names; Asked, allowing diverse answers; Excluded nonbinary respondents; Used data from professional societies \\
Watt et al. (2010) \cite{watt_adolescents_2019} & STEM-wide & First names (algorithm); Asked, M/F/prefer not to say only \\
Wilkinson et al. (2020) \cite{wilkinson_trends_2020} & Medical Sciences & First names (algorithm); Search text \\
Witteman et al. (2019) \cite{witteman_are_2019} & Medical Sciences & Asked, M/F only; Equated sex to gender \\
Wynn  \& Correll (2017) \cite{wynn_gendered_2017} & Technology & No methods provided \\
Wynn (2020) \cite{wynn_pathways_2020} & Technology & Asked, M/F only \\
Yoder \& Mattheis (2016) \cite{yoder_queer_2016} & STEM-wide & Asked, allowing diverse answers \\
Zeng et al. (2016) \cite{zeng_differences_2016} & STEM-wide & First names; Photos/videos \\
\label{table:1}
\end{longtable}
\end{landscape}

\subsection{Gender as White}

Researchers may analyze data on gender without analyzing data on race. This is a mistake, as gender and race are mutually co-constitutive. Gender is inseparably embedded in ethnic and racial contexts. For example, \say{bakla} is a gender identity in the Phillipines. Historically, bakla had a spiritual role there and currently face unique forms of oppression \cite{binaohan2014, Garcia_2000}. If someone identifies solely as bakla, it would be a misunderstanding to file them under \say{trans woman}, \say{gay man}, or some other ``non-racialized" gender category. However, even the concept of a non-racialized gender category is fallacious---the gender category examples given here are hegemonically white. They are the result of European colonization, which has violently dismantled gender epistemologies and replaced them with those of the colonizers \cite{johnston-2018-geography, lugones2007}.

Furthermore, the assumption that gender and race can be analyzed separately has led to poor scholarship and problematic policy demands \cite{crenshaw1989}. In the case of studies of STEM practitioners, we argue that separating gender and race will lead to mistaken beliefs about understanding and resolving gender disparities in science. Gender and race must be analyzed in conjunction to capture a complete understanding of how both impact the experience of STEM practitioners.

\subsection{Gender as Observable}

Many studies, explicitly or implicitly, employ a gender-by-name or gender-by-eye method to acquire gender data. This is a dubious, non-consensual, and flatly unethical approach which, simply put, would not pass peer review in a sociological journal. Often, the subjects of these studies are assigned a binary gender by their first names, either by the perception of the surveyor or by automated methods. If this method fails, many studies infer gender through real-time interactions or public records, such as photos or articles including third-person gendered pronouns (Table \ref{table:1}). Most often, this results in those with indeterminate gender simply being removed from the survey; this is particularly concerning when particular regions, countries, or cultures are specifically excluded from analysis, further emphasizing \say{Gender as White} \cite{huang_historical_2020_main}.

The treatment of gender as trivially discernible via name or presentation is unavoidably discriminatory. For people of marginalized genders, there is no acceptable outcome: we are either misclassified or  discarded. Misgendering and erasure are not trivial harms; they have very real psychological and professional consequences \cite{grant2011, mclemore2015,davidson2016,mizock2017,thorpe2017,cech_queer_2017_main,cech_lgbtq_2018_main}.

\subsection{Gender as Discrete}

A corollary to the above is that, in these studies, gender is treated as a discrete set of categories assumed to be stable and coherent across global populations, within individuals, and over time (Table \ref{table:1}). The male/female binary is usually employed, and while gender determination algorithms may not require gender to be binary, it does require it to be discretizable; e.g., through the selection of one of a limited number of options in surveys. Such frameworks unavoidably reduce people of marginalized genders to data points, losing nuance and denying subjects authority over how their gender is represented.  Assuming that gender is static over time also limits the ability for individuals to \say{come out} or otherwise affirm their gender across their lifetime. This erases binary transgender people as well as nonbinary and gender diverse people.

\subsection{Gender as a Statistic}

A common sentiment among these surveys is ``While we recognize that gender is not a binary, we do not include nonbinary/transgender/gender non-conforming people in our analysis due to a lack of statistical significance.'' This is an especially harmful attitude in the scheme of equity work. Statistical significance determines who gets to be accounted for---who counts. Reducing inclusion efforts to only those whose numbers are high enough to ``count'' is blatantly discriminatory and explicitly marginalizes a minority group. This produces results which are oversimplified at best and grossly inaccurate at worst. 

\subsection{Gender as Inconsequential}

A vast majority of studies described in Table \ref{table:1} focus on the experience of women in STEM. Some of them expand their definition of scope to include ``women and nonbinary people'' or ``women+'', terms which situate and dismiss nonbinary people as a sub-category of women. Related phrases such as ``female-identifying'' compound this effect by incorrectly implying that transgender women identify as, but are not, women. Even when inclusive language is used, spaces set aside for those who are not cisgender men often center cisgender women while sidelining other marginalized genders, enacting the same kind of gender-based marginalization they purport to subvert. This is a direct result of treating genders and experiences beyond those of cisgender men and women as inconsequential, superfluous, or unimportant. 

\subsection{Gender Work in STEM: Reinventing the Wheel}

Gender is not a new area of study, and gender inclusion is not a new problem. Gender studies and transgender studies \cite{Bhinder2021} are fields which have existed for decades. However, rather than deferring to established and well-researched paradigms of gender and its manifestations, STEM practitioners who publish the surveys described in this article choose to reinvent the wheel by ignoring the experts and their work. As a consequence, their results are rarely novel and often misinterpreted, causing harm to the communities they purport to serve. Furthermore, the self-survey of STEM practitioners incorrectly frames the problem of inclusion as one which can be solved without the input of the affected groups or established experts. 

\section{Suggested Best Practices and Recommendations}

\subsection{Methodological Choices}

Neither quantitative nor qualitative approaches to gender can come without a deep awareness of the complexity and context-dependence of gender. Any STEM practitioner who wishes to conduct gender-related research must begin by grounding the definition of gender they wish to investigate and aligning their data collection, analysis methods, and interpretation with said meaning. Quantitative methods should not be done away with; rather, a community-wide reckoning with the epistemic authority of these methods in matters of marginalization must occur. Furthermore, qualitative data and methods such as ethnographic description and participant testimony must be understood as inherently valuable. 

\subsection{Collecting and Reporting Gender Data}

There are two pillars to our suggested best practice for collecting and reporting of gender data: first, allow participants to self-report their genders, and second, never automate gender assignment.

Self-identification means that descriptions of participants' gender is provided by the participants themselves, rather than being assigned or assumed by researchers. Any question that does not provide a write-in option is inherently limited and risks excluding and erasing the genders of respondents whose identities were not anticipated by the research team. It is important to note that non-white people are typically harmed by this practice, as perceptions of legitimacy of gender are shaped by colonialism, white supremacy, and the cultural hegemony of the West.

For the collection of demographic data, we specifically recommend an open-text or fill-in-the-blank approach to gender data collection.

The wealth of information available online makes automation and analysis of “big data” a tempting path to creating larger and more powerful studies of gender in authorship, conference attendance, grant application, and more \cite{benjamens_gender_2020_main, shishkova_gender_2017_main, holman_researchers_2019_main}. However, these studies are almost guaranteed to misgender people within the dataset, even when the most stringent automation framework possible is used. Attempting to automate gender identification with machine vision or audio data is equally fraught. Gender is an internal quality; an individual is the sole authority on their gender identity \cite{nadal2023}. Automated assessments of external qualities such as name, appearance, or pronoun use can only draw valid conclusions about the qualities that they measure, and cannot reliably predict gender. Any study relying on automated gender assignation cannot produce valid results.

We emphatically discourage the gathering of gender data by any means other than voluntary self-identification, especially the usage of automated gender recognition algorithms. Journals and funding agencies should prioritize studies of marginalization in STEM which use the best practices described in this work and/or are conducted with the guidance of trained social scientists. For further reading on gender-inclusive data collection and survey design, we refer the reader to Rasmussen et al. (2019) \cite{rasmussen_nonbinary_2019}, Strauss et al. (2020) \cite{strauss_nonbinary_2020}, and references therein. 

\subsection{Privacy Considerations}

STEM communities range in size, and particularly small fields can be well-connected. Extreme caution must be exercised when preserving the privacy and anonymity of of survey participants, as these data alone can be used to identify marginalized individuals, especially those who are marginalized along multiple identities. If data need to be aggregated in order to maintain privacy of individuals, clearly lay out what aggregations were performed and why they were suitable for the specific study questions.

\subsection{Institutional Policies and Practices}

A full consideration of gender equity in the context of institutional reform is beyond the scope of this paper. At the core of equity work is the need for equity initiatives to recognize a more complex model of gender than previously employed. We recommend a shift from ``Women in STEM'' to ``Marginalized Genders in STEM''. This represents a major change to the status quo, but it is a necessary one as we move into a future where we support all members of marginalized communities. STEM curricula must reflect the human component of STEM: gender studies, transgender studies, critical race theory, and Science, Technology, and Society Studies. We recommend that study in these areas be added to undergraduate and graduate coursework and workforce development programs. 

\section{Conclusion}

Transgender, nonbinary, and gender non-conforming STEM practitioners have been failed by those who profess to support us. For many of us, this failure has defined, and ended, careers. Only by listening to and centering people of marginalized genders can we move forward in an equitable way to support every member of our STEM community. However, marginalized people are frequently called upon to educate others about the conditions of their marginalization, typically without compensation, and doing so may inadvertently disseminate and entrench the harm that nonbinary and gender non-conforming people experience throughout their careers in STEM. Before consulting marginalized people, we recommend that STEM practitioners wishing to educate themselves about gender first seek out resources independently to the best of their abilities, starting with the works cited in this paper.

The next tangible steps toward dismantling the unjust systems within which we live, and materially supporting the people hurt by these systems, will require sincere thought and work from both within and outside of our STEM community, not just passive claims of allyship or the completion of box-checking exercises to present a facade of inclusivity. These steps are required in order to repair the harm that has been perpetuated against our most vulnerable community members. The time to consider, to contemplate, or to form a committee has passed. By delaying action, a system that materially harms some of the most vulnerable members of our community is allowed to remain, dissuading the next generation of scientists and driving an exodus of talented researchers. The time to act is now.

\vspace{1cm}

\textbf{Acknowledgments:} We would like to thank S. B., A. E., B. J., D. J., E. S., and N. S. for their early discussions and contributions to this manuscript. We would also like to acknowledge our many transgender and nonbinary colleagues who have left STEM in the years since those initial discussions.

\newpage


\bibliography{sn-bibliography}

\section{Author Contributions}
Authors are listed in alphabetical order excluding the lead author and last author.


\end{document}